\documentclass[twocolumn,showpacs,superscriptaddress]{revtex4}
\usepackage{epsfig}
\usepackage{graphicx}
\usepackage{amsmath}
\usepackage{amsfonts}
\usepackage{upgreek}
\begin{document}

\preprint{APS/123-QED}

\title{It's harder to splash on soft solids}% Force line breaks with \\

\author{Christopher J. Howland}
\affiliation{%
Trinity College, University of Oxford, OX1 3BH, UK
}%
\author{Arnaud Antkowiak}
\affiliation{%
Institut Jean Le Rond dÕAlembert, UMR 7190 CNRS/UPMC, Sorbonne UniversitŽs, F-75005 Paris, France}%
\affiliation{%
Surface du Verre et Interfaces, UMR 125 CNRS/Saint-Gobain, F-93303 Aubervilliers, France
}%
\author{J. Rafael Castrej\'on-Pita}
\affiliation{%
School of Engineering and Materials Science, Queen Mary, University of London, E1 4NS, UK
}%
\author{Sam D. Howison}
\affiliation{%
Mathematical Institute, University of Oxford, OX2 6GG, UK
}%
\author{James M. Oliver}
\affiliation{%
Mathematical Institute, University of Oxford, OX2 6GG, UK
}%
\author{Robert W. Style}%
\email[]{robert.style@mat.ethz.ch}
\affiliation{%
Mathematical Institute, University of Oxford, OX2 6GG, UK
}
\affiliation{%
Department of Materials, ETH Z\"{u}rich, Z\"{u}rich 8093, Switzerland
}
\author{Alfonso A. Castrej\'on-Pita}%
\email[]{alfonso.castrejon-pita@wadh.ox.ac.uk}
\affiliation{%
Department of Engineering Science, University of Oxford, OX1 3PS, UK
}


%\affiliation{
%\affiliation{
%\affiliation{%

\date{\today}% It is always \today, today,

\begin{abstract}
Droplets splash when they impact dry, flat substrates above a critical velocity that depends on parameters such as droplet size, viscosity and air pressure.
By imaging ethanol drops impacting silicone gels of different stiffnesses we show that substrate stiffness also affects the splashing threshold.
Splashing is reduced or even eliminated: droplets on the softest substrates need over 70\% more kinetic energy to splash than they do on rigid substrates.
We show that this is due to energy losses caused by deformations of soft substrates during the first few microseconds of impact.
We find that solids with Young's moduli $\lesssim 100$kPa reduce splashing, in agreement with simple scaling arguments.
Thus materials like soft gels and elastomers can be used as simple coatings for effective splash prevention.
Soft substrates also serve as a useful system for testing splash-formation theories and sheet-ejection mechanisms, as they allow the characteristics of ejection sheets to be controlled independently of the bulk impact dynamics of droplets.
\end{abstract}

\pacs{Valid PACS appear here}% PACS, the Physics and Astronomy
\maketitle

%\tableofcontents

Splashing on solid surfaces is an active research topic with a wide range of applications, ranging from rainfall \cite{Cheng2014,zhao15}, to pesticide application \cite{knoc94}, inkjet printing \cite{Derby2010}, fuel combustion \cite{more10}, forensic science \cite{Bonn2014}, spray coating \cite{fauc04}, and the wide variety of systems where droplets impact liquid surfaces \cite{rein93,thor02,joss03,deeg07}.
However, recent key experiments have shown that the splashing process is more complex than previously thought so that there are many outstanding questions still remain regarding the physical mechanisms involved in droplet splashing \cite{ThoroddsenPRL2012, CastrejonVK2012, Xu2015}.
In this Letter, we study a new topic that has received little attention, namely how to eliminate splashing.
This is important for topics ranging from drop-based printing techniques to situations where splashing is undesirable due to hygiene or safety concerns (e.g. in accidental aerosolisation of toxic or biohazardous liquids upon spillage, or in maintaining sterile medical environments) \cite{harr04,davi07}.
Known splash-prevention techniques involve reducing air pressure \cite{xu05}, impacting elastic membranes \cite{pepp08b}, tilting the substrate \cite{liu10} or changing its speed \cite{bird09b}, and microscale surface patterning \cite{tsai10}.
However, none of these can be easily used to give all-round splash protection.
Here we show that soft coatings/substrates offer a novel solution that can significantly reduce and even eliminate splashing.
Beyond splash protection, our work is highly applicable to many situations in nature and technology where droplets impact upon soft surfaces like skin \cite{LohseSkin2013}, foodstuffs \cite{Osorio2011}, gels and emulsions.

We studied splashing on soft materials by impacting ethanol droplets on silicone/acrylic substrates with a large range of stiffnesses.
Silicone gels were made by combining a silicone base and crosslinker in different ratios and curing at room temperature \cite{styl15}. The resulting Young's moduli, $E$, were in the range $5-500$kPa, as measured by static indentation (they are approximately incompressible \cite{jeri11,styl13,styl14}).
Samples were 10mm thick unless otherwise stated.
Droplets were generated using a satellite-free droplet generator \cite{cast08,cast11} positioned at different heights above the sample, with which we were able to achieve a range of droplet sizes and impact velocities.
All experiments were in normal laboratory conditions at 22.5 $\pm$ 0.5$^\circ$C.
We visualised drop impacts with a shadowgraphy system consisting of a high intensity 100W white LED, an optical diffuser and a high speed camera (Phantom Miro310/V12.1).
We then extracted droplet sizes, velocities, and ejection dynamics from the videos with a bespoke Matlab programme.

\begin{figure}
\centering
  \includegraphics[width=9cm]{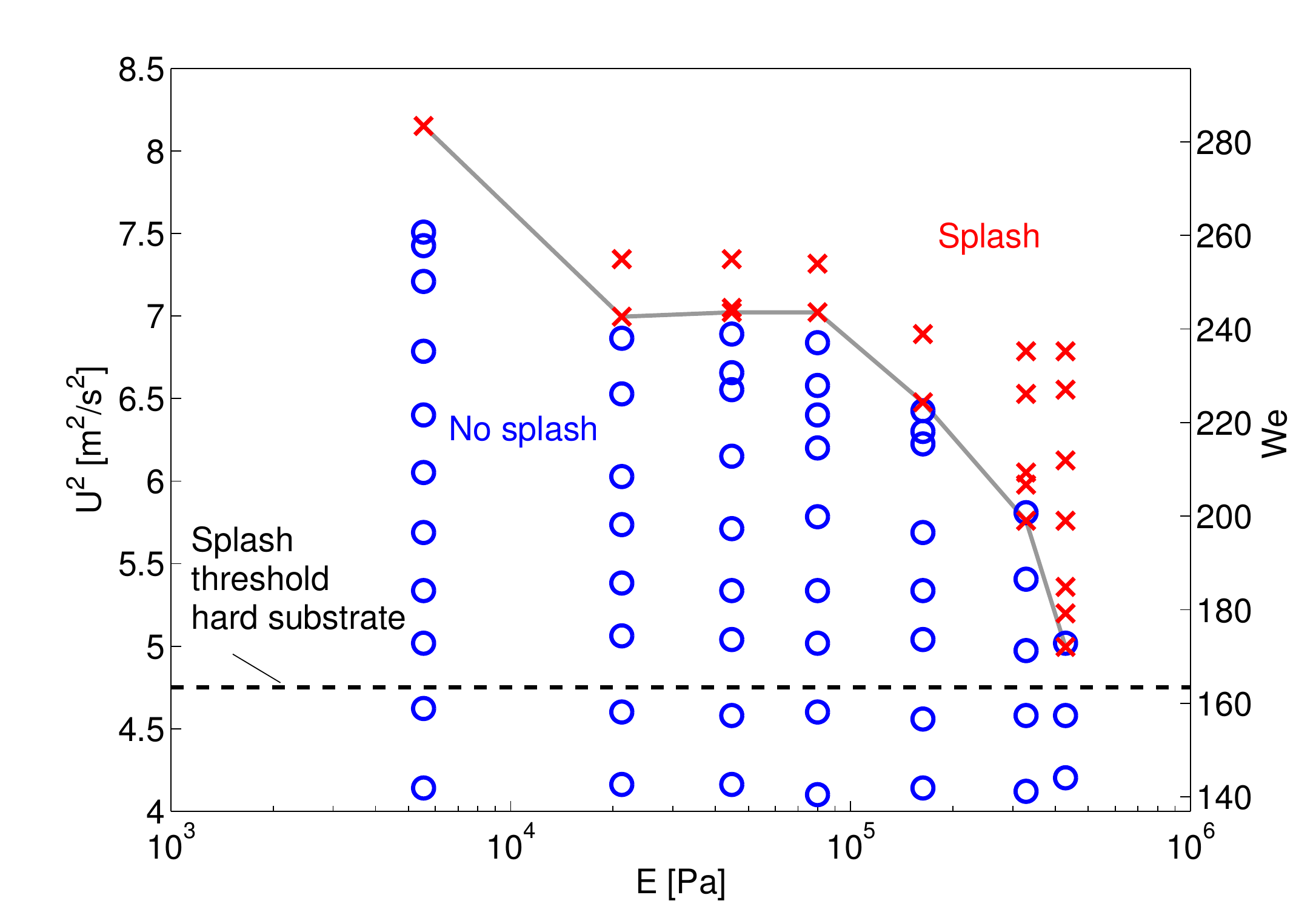}
  \caption{Splashing behavior of ethanol droplets on flat solid substrates as a function of Young's modulus and impact speed. All droplets have $R=0.97$mm, so $U^2$ is proportional to the Weber number (right hand axis). Circles: no splash, crosses: splash. The continuous curve indicates the splashing threshold (the lowest speed at which splashing was observed). This approaches the rigid-substrate splash threshold (dashed line) as $E$ increases.}
  \label{fig:splash_thresh}
\end{figure}

\begin{figure*}
\centering
  \includegraphics[width=15cm]{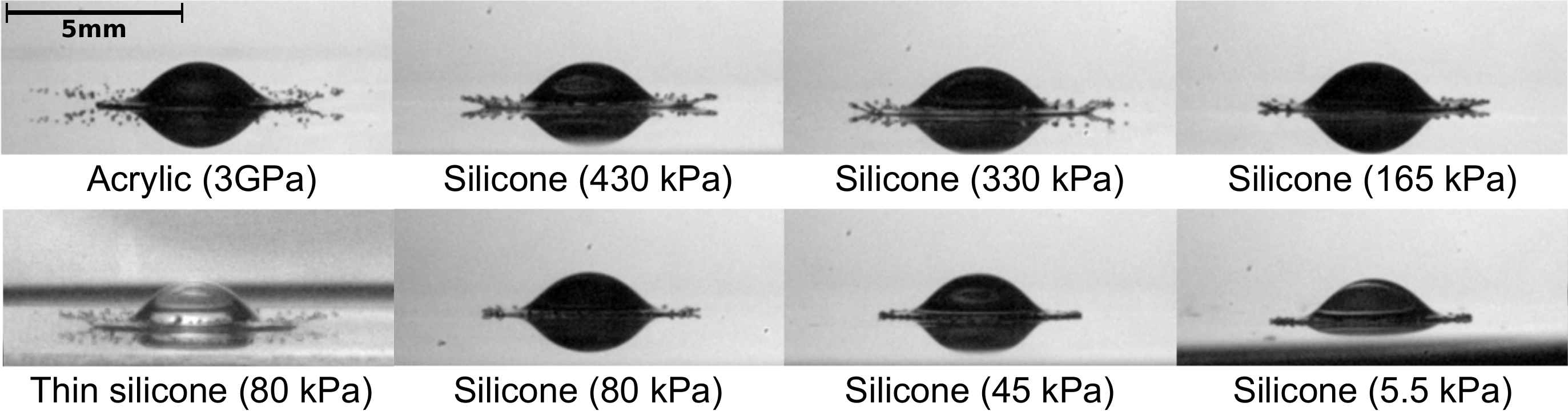}
  \caption{Examples of ethanol droplet impacts on flat substrates with a range of stiffnesses. All droplets have a radius of $0.88\pm 0.02$mm and impact speed of $2.61\pm 0.02$m/s. Images shown are taken approximately $350\mu$s after impact. All silicone substrates are 10mm thick, except the bottom left image which is $3\mu$m thick. The scale bar is the same for all eight images.}
  \label{fig:ex_impacts}
\end{figure*}

Splashing is significantly reduced and even eliminated on soft surfaces.
Initially, we impacted droplets on two types of samples: rigid, flat acrylic samples ($E\sim 3 $GPa) and flat silicone gels cured in 90mm diameter, 10mm deep moulds.
All droplets had radii $R= 0.97 \pm 0.07$mm and impacted at speeds $U$ from 2--3m/s.
Note that substrate wettability should not affect impact behavior in this range \cite{pasa96}.
The droplets behave qualitatively the same as previously-observed high Reynolds number (Re) impacts (e.g. \cite{riob01,yari06,ribo14} and Supplementary Videos).
Shortly after impact, a thin sheet is ejected radially outwards from near the droplet's apparent contact line.
If the sheet is ejected at high enough velocity, it lifts up from the substrate and subsequently breaks up into droplets (a corona splash) \cite{ribo14}.
At lower ejection speeds the sheet initially lifts away from the substrate, but it falls back down onto the substrate where it rapidly slows before splashing can occur.
Figure \ref{fig:splash_thresh} shows at which speeds or, equivalently, Weber number ($\mathrm{We}=\rho U^2 R/\gamma$, $\rho$ and $\gamma$ are droplet density and surface tension respectively) splashing occurs as a function of substrate stiffness.
For any given substrate, there is a threshold velocity for splashing, which increases with increasing substrate compliance.
The rigid (acrylic) splash threshold, $U=2.18$m/s, is shown by the dashed line in the figure.
The splashing behavior on the stiffest silicone substrates approaches this limit.

Figure \ref{fig:ex_impacts} illustrates impact behavior.
All images show droplets of the same size which have hit different substrates at the same speed (2.61 $\pm$ 0.02m/s) approximately $t=350\mu$s after impact.
For the 10mm thick samples (all except the bottom left-hand image) there is a smooth transition from violent splashing to no splashing with decreasing $E$.
The position of the leading edge of the ejection sheet shows that ejection is significantly faster on stiffer substrates than on softer ones.
Videos of the droplet impacts on acrylic and silicone with $E=165$ and $45$ kPa are given in the Supplementary Material.

The reduction in splashing on soft substrates is due to deformations caused during the droplet impact process.
We showed this by impacting droplets on a $3\mu$m thick coating of a soft gel ($E=80$kPa) spincoated onto a glass slide (this limits deformations to $O(\mu\mathrm{m})$).
Impacts on this surface were almost identical to impacts on acrylic surfaces (with the same splash threshold velocity of $U=2.18$m/s) and much more violent than impacts on a deep substrate made from the same silicone (e.g. Figure \ref{fig:ex_impacts}).
This also rules out the splash reduction being caused by changes in the surface properties of silicone with $E$.

\begin{figure*}
\centering
  \includegraphics[width=17.5cm]{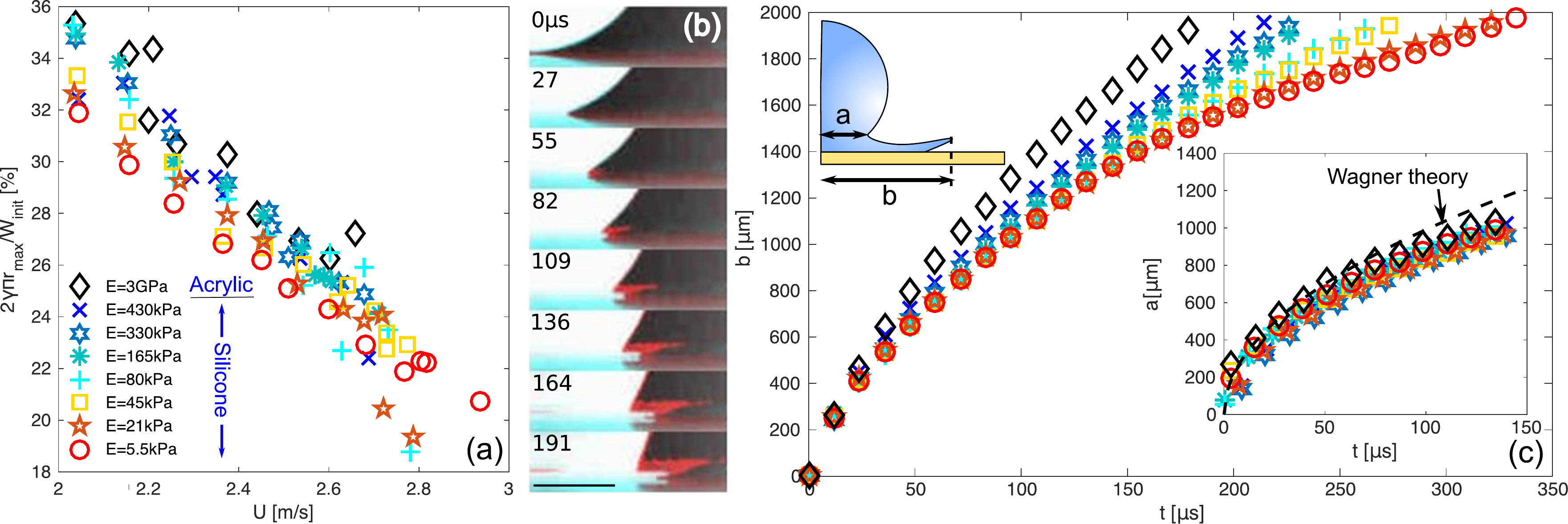}
  \caption{(Color online) a) The estimated percentage of droplet energy remaining after impact and spreading for the experiments in Figure \ref{fig:splash_thresh}. Soft substrates only absorb a few percent more energy than the rigid substrates. This can not explain why splashing on soft substrates requires 75\% more energy than on rigid ones. b) Overlaid images of the initial stages of sheet ejection during ethanol droplet impact on rigid (acrylic) and soft (silicone with $E=5.5$kPa) substrates. $U=2.36$m/s and $R=1.5$mm. Red/cyan corresponds to rigid/soft substrates. The image appears white when the position of the two droplets coincide. Sheet ejection is clearly much faster on the harder substrate, while the bulk droplets deform very similarly. The scale bar is 1mm. c) The radial position of the ejection-sheet tip (main figure) and the turnover point (inset) for droplets with $U=2.62$m/s and $R=1.25$mm. Ejection-sheet velocities smoothly reduce as $E$ reduces. The motion of the turnover point is relatively independent of stiffness and agrees quite well with the Wagner theory prediction (dashed curve), $a=\sqrt{3 R U t}$.}
  \label{fig:energy}
\end{figure*}

Although substrate deformations absorb energy from an incoming droplet \cite{mang12}, they do not merely absorb  the relatively large energy excess required to splash on a soft surface.
The pre-impact energy of a droplet is $W_{init}=2 \pi R^3 \rho U^2/3+4\pi R^2 \gamma$, with the majority of this being kinetic energy.
Thus, splashing on the softest substrate can require over 75\% higher $W_{init}$ than on rigid substrates (Figure \ref{fig:splash_thresh}).
We can estimate the energy absorbed by the substrate during impact and spreading by calculating the surface energy of the droplet at its maximum spread radius $R_{max}$.
At this point, the kinetic energy is approximately zero (the energy in ejected microdroplets is minimal), so the droplet energy $\sim 2 \gamma \pi R_{max}^2$ \cite{clan04}.
Figure \ref{fig:energy}(a) shows this, normalised by $W_{init}$.
For each droplet impact speed, the energy dissipation on soft substrates and hard substrates are only a few percent apart.
This is nowhere near enough to explain the observed splashing reduction, and suggests that the substrate stiffness has relatively little effect on the kinetics of the bulk of the droplet.

Instead, the splash-suppression mechanism predominantly removes energy from the very small volume of liquid that forms the ejecta.
The resulting ejection sheet is thus slower, less energetic, and cannot break up into a splash as easily. 
For example, Figure \ref{fig:energy}(b) shows overlaid pictures of droplets impacting hard (acrylic, red) and soft ($E=5.5$kPa silicone, cyan) substrates filmed at 110000 fps.
For both droplets, $U=2.36$m/s and $R=1.5$mm.
The motion of the bulk part of the droplets is relatively unaffected by substrate stiffness.
However, sheet ejection is clearly more violent on the harder surface; it travels faster, rising up off the substrate and subsequently breaking up into droplets, in contrast to the sheet on the softer substrate.
This is shown quantitatively in Figure \ref{fig:energy}(c), which gives the ejection-sheet tip position, $b(t,E)$, and the turnover point, $a(t,E)$ (cf schematic).
Pre-ejection we take $a$ and $b$ to be the apparent contact line of the droplet.
The ejection-sheet velocity increases significantly with $E$: at late times, $t$, spreading is more than twice as fast on the rigid substrate as it is on the softest substrate.
However the growth of $a$ is apparently independent of $E$.

To understand how splashing changes on soft substrates, we need to understand sheet ejection, and how it changes on soft surfaces.
Unfortunately this occurs at very small time- and length-scales, so it is difficult to observe experimentally \cite{pepp08b,mand09,mand12,koli12}.
However we can use simulations, and inviscid theory \cite{howi91,oliv02,scol03} to understand what occurs.

We perform numerical simulations with Basilisk software \cite{popi,popi09,phil16}, modelling 3d impacts onto simple substrates that respond to overlying pressures with a simple viscoelastic response: $p'=ky/R+\eta \dot{y}/U$, where $p'(x,t)=p/\rho U^2$ is the nondimensional pressure immediately above the surface, $y(x,t)$ is the height of the substrate and $k,\eta$ are constants.
The simulations include air, surface tension, and viscosity (full details in the Supplement), and we take $\mathrm{We}=150$, $\mathrm{Re}\equiv \rho U R/\mu=2500$, a liquid/air density ratio of $\rho/\rho_a=82$ and a liquid/air dynamic viscosity ratio $\mu/\mu_a=5.6$ (to allow for a reasonable computational time).

The results have the same behavior as our experiments.
The bulk of the droplet behaves practically identically for all the substrates, while as the substrate stiffness reduces, the sheet is ejected significantly slower despite only small substrate deformations occurring (see Supplement).

\begin{figure}
\centering
  \includegraphics[width=8cm]{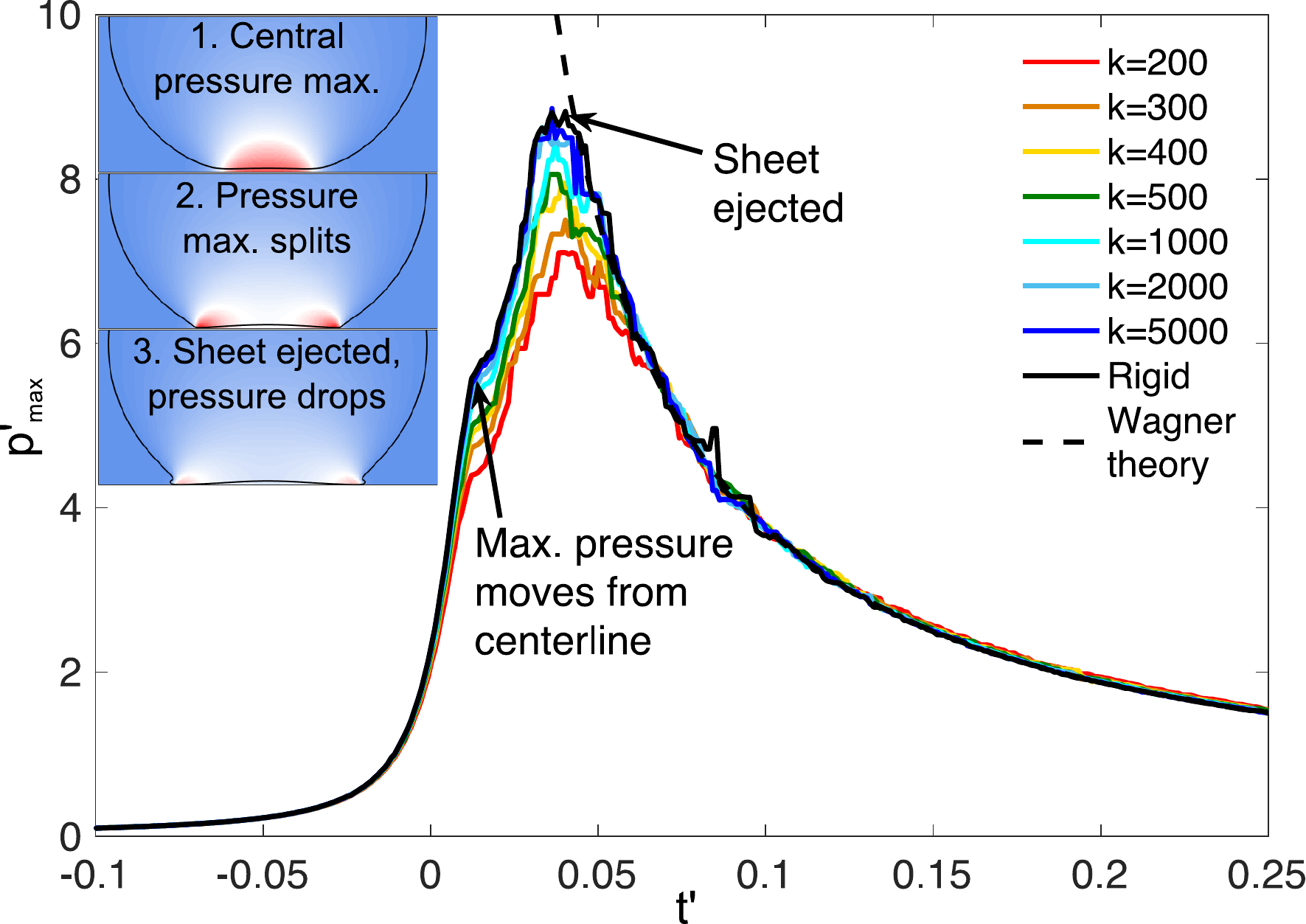}
  \caption{(Color online) Simulations give the maximum pressure $p'_{max}\equiv p_{max}/\rho U^2$ exerted on a substrate as a function of time $t'\equiv(t-t_i)U/R$. The insets illustrate how the pressure distribution evolves in an impacting droplet (here impacting a rigid substrate). $p'_{max}$ reaches a peak at the point of sheet ejection before decaying following Wagner theory's prediction $p'_{max}=3/8t'$ (dashed curve).}
  \label{fig:energy}
\end{figure}

The numerical simulations show a common sequence of events upon impact (see snapshots in Figure \ref{fig:energy}(c)).
First, as the droplet starts to approach the substrate, it slows down due to the presence of air trapped ahead of it.
This deceleration causes a pressure maximum under the centerline of the droplet, which increases as the contact patch of the droplet spreads out, trapping a thinning air pocket.
Second, soon the position of the maximum pressure in the droplet moves away from the centerline, following the advancing `contact line' of the droplet \cite{howi91,mand09,phil16,egge10}.
At the same time, the separating air film ceases to thin, as the pressure gradient at the droplet base now drives air towards the centerline of the droplet.
Finally, the pressure in the droplet reaches a maximum as the sheet is ejected from its base.

The key control for splashing is the maximum pressure, $p_{max}(t,E)$, on the surface of the substrate: this both provides the driving force for sheet ejection, and can cause the largest substrate motions (recent work predicts splashing when the sheet-ejection speed $V_e>C \gamma/\mu_a$, with $C$ a constant \cite{ribo14}).
Figure \ref{fig:energy}(c) shows typical $p_{max}$ traces for a variety of different substrates (here $k/\eta=10$).
Each plot clearly shows the three different stages described above.
Interestingly, the late time behavior after sheet ejection always shows excellent agreement with $p_{max}=3\rho RU/8(t-t_i)$, an expression which follows from Wagner theory's asymptotic results that $p_{max}=\rho \dot{a}^2/2$ \cite{howi91} and $a=\sqrt{3 R U (t-t_i)}$ \cite{ribo14} (see also the good agreement in Figure \ref{fig:energy}(c)).
Here $t_i$ is the `impact time' that is calculated by extrapolating the drop's position before it is slowed by air-cushioning.
Importantly the peak pressure, $p_{peak}$, occurs at the ejection time, $t_{ej}$, at the start of the `Wagner regime'.
Thus $p_{peak}=3RU/8(t_{ej}-t_i)$, a result that also allows us to measure $p_{peak}$ from experimental observations.
Note that Wagner theory ignores air, surface tension and substrate rheology, so its failure to describe $p_{max}$ at early times indicates that these must play a dominant role then.

The simulations also allow us to understand why $p_{peak}$ is lower (and thus splashing is less likely) on softer substrates.
We find that a reduction in $p_{peak}$ is associated with downwards substrate motion, as this reduces liquid deceleration on the substrate.
Significant substrate motion does not occur until when $p_{peak}$ exceeds the substrate's modulus, $E$.
Thus splash reduction by a soft substrate can only occur when
\begin{equation}
p^r_{peak}=\frac{\rho RU}{(t^{r}_{ej}-t^r_i)}\gtrsim E.
\end{equation}
Here superscript $r$'s refer to impact on a rigid substrate.
This result agrees with our experimental results.
Using the data from Figure \ref{fig:energy}(b), we take $R=1.5$mm, $U_c=2.36$m/s, $t^{r}_{ej}-t^r_i=30\mu$s and $\rho=789\mathrm{kg}/\mathrm{m}^3$, to find $E_c=93$kPa.
This is consistent with substrate stiffnesses where we start to observe significant reductions in splashing.

To predict how the splashing threshold changes on soft substrates, as seen in Figure \ref{fig:splash_thresh}, is more complex.
For this, one will require a quantitative model for how the ejection-sheet velocity and thickness depend on parameters like $R,$ $U$, $p_{peak}$ and the frequency-dependent substrate rheology.
The onset of splashing then needs to be related to the ejection-sheet characteristics through one of several competing theories of splashing \cite{mund95,ribo14,liu15,joss16}.
For example, as mentioned above, our proposed mechanism is consistent with a recent proposal that, for splashing, sheet ejection needs to exceed a critical speed to overcome capillary retraction \cite{ribo14}.
A second recent proposal is that splashing occurs when the ejection-sheet thickness, $d=3\gamma\sqrt{2\pi \kappa}/2 \rho_a \dot{b} c$ due to a resonance driven by the Kelvin-Helmholtz instability \cite{xu07,liu15}. Here $c$ and $\kappa$ are the speed of sound in air, and the adiabatic gas constant of the air respectively.
It is difficult to measure $d$ precisely enough to test this mechanism.
However, for typical sheet velocities from Figure \ref{fig:energy}(c), this result predicts that ejection sheets will break up over acrylic substrates when they are $O(15\mu$m) thick, and approximately double this thickness on the softest substrate.
These values are comparable to sheet observations in previous studies \cite{liu15}, so this mechanism also seems consistent with our results.

In conclusion, droplet splashing is reduced and can even be eliminated on soft surfaces.
Soft substrates do not affect the bulk behavior of an impacting droplet.
However, small, early-time substrate deformations can absorb much of the energy from an ejection sheet, preventing it from breaking up and forming a splash.
This allows droplets to have significantly higher splash-free impact speeds on soft surfaces than on hard ones: in our experiments, the minimum fall height for splashing almost doubled from the rigid surface to the softest substrate (from 29cm to 55cm).
By considering the pressure build up under an impacting droplet, we have shown that surfaces can only cushion splashing when $E\lesssim \rho R U /(t^r_{ej}-t^r_i)$.
For typical millimetric falling droplets, this corresponds to substrates with $E\lesssim O(100)$kPa, meaning that a wide range of gels and elastomers can be used as a way of fabricating novel, inexpensive, splash-proof coatings.
Soft substrates are also a new system for investigating the detailed mechanics of droplet impact.
For example, they can be used to test different splash-formation mechanisms, as they allow fine control of ejecta characteristics independent of the motion of the bulk droplet.
Additionally, by patterning substrates with soft/stiff or shallow/deep areas (e.g. \cite{styl13b}) one can control the pressure distribution in a droplet upon impact, allowing detailed experiments on ejecta mechanics.
Finally, our work should give insight into the wide variety of processes involving impact on soft substrates, with examples ranging from maintaining hygiene, to pesticide delivery on plants, optimising inkjet printing on soft/biological materials and circuit printing of soft electronics.

The authors gratefully acknowledge helpful conversations with D. Vella, A. Goriely, C. MacMinn, E. Dufresne and K. Jensen. CH was supported by an EPSRC Vacation Bursary. RWS and AACP received funding from the John Fell Oxford University Press (OUP) Research Fund. AACP was funded by the Royal Society via a University Research Fellowship and a Research Grant.

\end{document}